\documentclass{elsart}   % Long art.

\usepackage{graphicx}
\usepackage{epsfig}
\usepackage{bm}
\usepackage{latexsym}

\begin{document}

%----------------------------------------------------------------------
%                               TITLE AND AUTHORS
%----------------------------------------------------------------------
\begin{frontmatter}
  
  \title{Diffusion and networks:\\A powerful combination!}
  \author{Ingve Simonsen}
                \ead{Ingve.Simonsen@phys.ntnu.no} 
  \address{Department of Physics\\ 
    %NTNU,
    Norwegian University of Science and Technology\\
 NO-7491 Trondheim,
    Norway}
  \date{\today} 
  \maketitle

%----------------------------------------------------------------------
%  ABSTRACT AND KEYWORDS
%----------------------------------------------------------------------

\begin{abstract}
  Over the last decade, an enormous interest and activity in complex
  networks have been witnessed within the physics community. On the
  other hand, diffusion and its theory, have equipped the toolbox of
  the physicist for decades.  In this paper, we will demonstrate how
  to combine these two seemingly different topics in a fruitful
  manner.  In particular, we will review and develop further, an
  auxiliary diffusive process on weighted networks that represents a
  powerful concept and tool for studying network (community)
  structures. The working principle of the method is the observation
  that the relaxation of the diffusive process towards the stationary
  state is {\em non-local} and fastest in the highly connected regions
  of the network. This can be used to acquire non-trivial information
  about the structure of clustered and non-clustered networks.
\end{abstract}

% Keyword

\end{frontmatter}

%--------------------------------------------------------------------
%   MAIN TEXT
%--------------------------------------------------------------------

\section{Introduction}
\label{Sect:Introdution}

Diffusion processes arise very naturally in a number of physical,
chemical and engineering problems. The topic has, therefore, attracted
a lot of attention by numerous brilliant scientists for more than a
century.  Early pioneers of the field were well-known scientists like
Einstein, Smoluchowski, Langevin, Wiener, Ornstein, Uhlenbeck {\em
  etc.} This year, in fact, we celebrate the one hundred year
anniversary of Einstein's seminal 1905 paper on the kinetic theory of
Brownian motion~\cite{Einstein1905,Einstein1956}.  To acknowledge
this event, as well as the other two influential ground breaking
papers by Einstein from the same year,
United Nations has appointed year 2005 the {\em World Year of Physics}.
So what can be more appropriate than choosing the title {\em Diffusion
  and soft matter physics} for this years {\em Karpacz Winter School of
  Theoretical Physics}.

Today there exists a well developed theory of
diffusion~\cite{DiffusionTheory} --- a research field that still is
vibrant and very much alive. The theory is capable of successfully
describing a number of natural occurring processes. However,
diffusion, and the concept of random walks, first introduced by
Perrin, are also useful concepts outside the branch of natural
processes.  This very paper might serve as one particular (out of
many) example of such. Herein we will apply diffusion as a concept, or
tool, to study a problem that has no direct connection to diffusion.
In particular, what will be considered is the (large scale) topology
of networks.

Complex networks are abundant in nature and society. They are set of
objects with some relations defined among them, resulting in
complicated non-regular structures.  A prototype example, taken from
sociology, is a group of people (the objects) where social
acquaintances represent the relations (known as edges or links)
between the objects.  The readers unfamiliar with networks are
encouraged to consult Refs.~\cite{Albert2002,Newman2003} for a general
introduction to the topic as well as numerous examples of real-wold
networks.

Traditionally the topology of networks has been studied by visual
inspections.  This was made possible since the number of objects,
known as vertices or nodes, was typically rather small.  However, with
the advent of the computer and an increased use of networks in
technological applications, the size of the studied networks started
to grow rapidly.  Today, like in, say, internet and web-page networks,
the number of nodes can reach millions or more.  Under such
circumstances, visual analyzing tools are not appropriate, and new
methods for their study are needed. It was at this point in time in
the history of network analysis that the method of statistical
physics, and the physicists that know them, entered the scene.

The present paper will, in the spirit of the winter school, combine
diffusion with a topic from soft matter physics --- complex networks.
In particular, what will be done is to report on, and extend, previous
works~\cite{Eriksen_PRL,Simonsen-2003} where an auxiliary random walk
process was used to characterize large topological features of complex
networks. Of special interest is the ability to locate and identify
community structures, a topic that has attracted a great deal of
attention lately~\cite{Girvan2002,Newman2004,Donetti2004}. Network
clusters, or community structures, are characterized by a subset of
vertices of the network having a considerably larger number of edges
among themselves than to vertices outside the subset.  In such cases
the subset is said to form a network community (or cluster).
 
Recently there has been quite some interest in the study if weighted
networks~\cite{Almaas2004,Newman2004-A}. To incorporate the weight of
edges into the analysis of network can be critical for determining,
say, its structure.  However, it is only recently that such studies
have been taken up upon by be the community in general. In this paper
we will incorporate the weights of the edges into the diffusion, or
random walk, formalism that was developed
previously~\cite{Eriksen_PRL,Simonsen-2003}. Herein we will review and
extend the presently known results to weighted networks.

This paper is organized as follows: In Sec.~\ref{Sect:MatsterEq}, the
foundation of the diffusion approached is derived, that is, the master
equation and its solutions. Then we address the so-called {\em current
  mapping technique} that utilize these solutions in order to uncover
information about the large scale topology of networks
(Sec.~\ref{Sect:CurrentMappingTech}).  The application of this
technique to various types and sizes of real-world networks is
presented in Sec.~\ref{Sect:Results}. We finally round off the paper
in Sec.~\ref{Sect:Conclusions} by presenting the conclusions.

\section{The master equation}
\label{Sect:MatsterEq}

Consider a network consisting of a set of vertices (of one single
type) and {\em weighted}, directed edges connecting them. It will be
assumed, for simplicity, that the network represents a single
component, {\it i.e.}  any pair of vertices can be reached by
following the edges of the graph. The weight associated with the edge
from, say, vertex $j$ to $i$, will be denoted $W_{ij}$ and
corresponding to the elements of the weighted adjacency matrix.

We will study diffusion (or random walks) on such networks and derive
the {\em master equation} that governs the time development of the
process.  The derivation parallels the one given previously for
unweighted, undirected networks \cite{Eriksen_PRL,Simonsen-2003}.
One starts by imagining placing a large number of (random) walkers
onto the vertices of the network.  These walkers are allowed, in each
time step, to move between adjacent vertices along the directed edges
connecting them. 
What edge, out of the possible (outgoing) ones, a walker chooses to
move along, is picked randomly with a probability that is {\em
  proportional} to the weight associated with that (directed) edge.
The different outgoing edges leaving a given vertex will therefore in
general, unlike the unweighted case~\cite{Eriksen_PRL,Simonsen-2003},
have different probabilities for ``accepting'' walkers. In this way
the system evolves in time.

Let the number of walkers ``living'' on vertex $i$ at time $t$ be
$N_i(t)$. Then the fraction of walkers at this vertex, out of a total
of $N$, is $\rho_i(t)=N_i(t)/N$.  The starting point of the derivation
of the master equation that describes the walker dynamics on the
network, is the observation that the total number of walkers is
guaranteed to be constant at all time, {\it i.e.}  $\sum_i \rho_i(t) =
1$ for {\em every} $t$.  Furthermore, the change in the walker density
of a vertex $i$ during {\em one} time step, equals the difference
between the relative number of walkers {\em entering} and {\em
  leaving} the same vertex over the time interval.  In mathematical
terms one may write\footnote{This equation resembles the continuity
  equation of, say, diffusing particles : $\partial_t\rho+\nabla
  \!\cdot\!{\bm J}=0$.}
\begin{eqnarray}
  \label{eq:ME2}
  \rho_i(t+1)&=&\rho_i(t) + J_i^{(-)}(t)-J_i^{(+)}(t),
\end{eqnarray}
where $J_i^{(\pm)}(t)$ denote the relative number of walkers
entering~($-$) and leaving~($+$) vertex $i$. How many walkers that
leaves along the different outgoing edges of vertex $i$ depends on the
total outgoing weight of this vertex, $\sum_k W_{ki}$. The fraction of
outgoing walkers from vertex $i$ (a current) {\em per} unit weight, is
thus
\begin{eqnarray}
  \label{eq:ME4}
  c_i(t) &=& \frac{\rho_i(t)}{\sum_k W_{ki}},
\end{eqnarray}
so that the edge {\em current} on the directed edge from vertex $j$
towards $i$, is given by\footnote{The magnitudes of these currents
  measure how important a link is. They are therefore intimately related
  to the edge betweenness, so that a high value of this latter
  quantity corresponds to a high value for the edge current.}
\begin{eqnarray}
  \label{eq:5}
  C_{ij}(t) &=& W_{ij}\,c_j(t) \;=\;  W_{ij}\frac{\rho_j(t)}{\sum_k W_{kj}}. 
\end{eqnarray}
Notice that the factor $W_{ij}/\sum_k W_{kj}$ is the probability of a
walker deciding on the edge from vertex $j$ to $i$. By adding {\em
  all} outgoing edge currents from vertex $j$, the relative number of
outgoing walkers (from $j$) will result; $J_j^{(+)}(t)=\sum_i
C_{ij}(t)$.  Substituting Eq.~(\ref{eq:5}) into this expression, one
readily demonstrates that $J_j^{(+)}(t)=\rho_j(t)$. This expresses the
fact that all walkers at vertex $j$ at time $t$, will leave it in the
next time step.  Similarly, one finds for the walkers leaving vertex
$j$, $J_j^{(-)}(t)=\sum_i C_{ji}(t)$, but now the expression can not
be simplified further.  Introducing the expressions for
$J_i^{(\pm)}(t)$ into Eq.~(\ref{eq:ME2}) results in:
\begin{eqnarray}
  \label{eq:master-eq}
   \partial_t \rho_i(t)
&=&
    \sum_j T_{ij} \rho_j(t) - \rho_i(t),
\end{eqnarray}
where $\partial_t \rho_i(t) = \rho_i(t+1)- \rho_i(t)$ and 
\begin{eqnarray}
  \label{eq:transfer}
  T_{ij} &=& \frac{W_{ij}}{\sum_k W_{kj}}. 
\end{eqnarray}
Moreover, this equation can easily be casted into the following matrix
form
\begin{eqnarray}
  \label{eq:ME5}
  \partial_t{\bm \rho}(t) &=& {\bm D}{\bm \rho}(t),
\end{eqnarray}
where $D_{ij}= T_{ij} - \delta_{ij}$, and it is the earlier announced
{\em master equation} for the random walk dynamics on the network.  It
resembles the diffusion equation, so we have termed ${\bm D}$ the
diffusion matrix (or operator).  Alternatively, Eq.~(\ref{eq:ME5})
could be reformulated as
\begin{eqnarray}
  \label{eq:ME7}
   {\bm \rho}(t+1) &=& {\bm T}{\bm \rho}(t),
\end{eqnarray}
where the elements of ${\bm T}$ are defined by
Eq.~(\ref{eq:transfer}). Notice that Eqs.~(\ref{eq:ME5}) and
(\ref{eq:ME7}) are in principle equivalent.  Physically,
Eq.~(\ref{eq:ME7}) means that ${\bm T}$ transfers (propagates) the
walker density ${\bm \rho}(t)$ one step forward in time. Due to this
property, ${\bm T}$ has been termed the {\em transfer
  matrix}~\cite{Eriksen_PRL,Simonsen-2003}. The attentive reader
should check, and find, that in the special case of an unweighted
network, {\it i.e.}  $W_{ij}=A_{ij}$ with $A_{ij}$ being the
unweighted adjacency matrix, Eqs.~(\ref{eq:ME5}) and (\ref{eq:ME7})
reduce to the expressions that were reported previously in
Refs.~\cite{Eriksen_PRL,Simonsen-2003}.  

It is often of advantage to work directly with the currents (per unit
edge weight) ${\bm c}(t)$ instead of the walker densities ${\bm
  \rho}(t)$. An equation satisfied by these currents can be obtained
from Eq.~(\ref{eq:ME7}) by dividing it through by $\sum_k W_{ki}$.
After recalling Eq.~(\ref{eq:ME4}), it is straightforward to arrive at
\begin{eqnarray}
  \label{eq:master-eq-vector}
  {\bm c}(t+1) &=& {\bm T}^\dagger {\bm c}(t),
\end{eqnarray}
where ${\bm T}^\dagger$ denotes the adjoint of ${\bm T}$. Thus,
technically ${\bm T}^\dagger$ is the transfer matrix for the currents
${\bm c}(t)$. In a similar way, the adjoint of the diffusion matrix
will play the role for the currents that ${\bm D}$ did for the walker
densities\footnote{In order to show this, simply add ${\bm c}(t)$ to
  both sides of Eq.~(\protect\ref{eq:master-eq-vector}).}.
The governing equations for the currents ${\bm c}(t)$ are thus
analogous to Eqs.~(\ref{eq:ME5}) and (\ref{eq:ME7}) accept for the use
of the adjoint matrices.

We will now demonstrate that the master equation supports a stationary
solution, {\it i.e.} a solution that does not depend on time. The
easiest way to show this is to start from Eq.~(\ref{eq:ME7}) and
conjecture that the stationary state satisfies:
$\rho_i(\infty)\propto\sum_j W_{ji}$.  This form is motivated by what
was previously found for unweighted
networks~\cite{Eriksen_PRL,Simonsen-2003} where in the stationary
state the walker density of a vertex is proportional to its degree.
By introducing this expression for $\rho_i(\infty)$ into
Eq.~(\ref{eq:ME7}) and recalling Eq.~(\ref{eq:transfer}), one readily
finds that $\rho_i(\infty)$ indeed is a stationary state, but only if
$\sum_j W_{ij}=\sum_jW_{ji}$ for all $i$'s. This implies that a
stationary state exists if the total {\em outgoing} and {\em incoming}
weight of each vertex of the network are equal. Notice, that this is
trivially satisfied for an undirected network, but also a sub-class of
directed graphs satisfies this requirement. In the stationary state,
the walker densities are therefore proportional to the total outgoing
weight ($w_i=\sum_j W_{ji}$) of the vertex, and hence according to
Eq.~(\ref{eq:ME4}) the current per unit outgoing weight will just be
constant; ${\bf c}(\infty)\propto 1$.

Formally the stationary state corresponds to the unit eigenvalue of
${\bm T}$ (or ${\bm T^\dagger}$) that turns out to also be the largest
possible eigenvalue~\cite{Eriksen_PRL,Simonsen-2003}.  In fact it is
of interest to know a number of the largest eigenvalues and the
corresponding eigenvectors of ${\bm T}$(or ${\bm D}$). The reason
being, as was explained in detail in Ref.~\cite{Simonsen-2003}, that
they control the relaxation towards the stationary state of the
slowest decaying modes of the diffusive process on the network.  It
should be mentioned, that one can show, like for the case of
unweighted networks, that the non-symmetric matrix, say, ${\bm T}$, is
similar to the symmetric matrix ${\bm K} {\bm T}{\bm K}^{-1}$ where
$K_{ij}=\delta_{ij}/\sqrt{w_i}$ and $w_i=\sum_jW_{ji}$.  Hence, ${\bm
  T}$ is guaranteed to have {\em real} eigenvalues and
eigenvectors~\cite{Meyer-2000}. It is practical (and usual) to sort
the real eigenvalues so that $\lambda^{(1)}$ corresponds to the
largest eigenvalue of ${\bm T}$, $\lambda^{(2)}$ the next to largest,
and so on. Below we will silently assume that this convention is
followed and collectively denote the eigenvalues by
$\lambda^{(\alpha)}$ where $\alpha=1,2,\ldots$ is the {\em mode}
index.  Moreover, all eigenvalues of ${\bm T}$ fall in the range
$-1<\lambda^{(\alpha)}\leq 1$, as is a consequence of the number of
walkers being conserved at all time.  The largest eigenvalue
$\lambda^{(1)}=1$ will, as a consequence of the
Perron-Frobenius theory (non-negative matrices)~\cite{Meyer-2000}, be
unique for a single component network and the elements of the
corresponding eigenvector will all have the same signs.

% \subsection{Numerical implementation}

% For the diffusion approach to be practically useful for large
% real-world networks, a fast, memory saving, and optimized algorithm
% for the calculation of the largest eigenvalues and the corresponding
% eigenvectors of a sparse matrix is required. The most time consuming
% part of the method is the eigenvector calculation. One such fast
% algorithm is due to Lanczos~\cite{Lanczos,Book:Golub}. It is
% implemented and made available via for instance the TRLan software
% package~\cite{TRLan}.  This software is optimized for handling large
% problems, and it can run on large scale parallel supercomputers.

\section{The current mapping technique}
\label{Sect:CurrentMappingTech}

Part of the power of the network diffusion approach lies in the
current mapping (or projection) technique. It is based on the
observation that vertices being connected to each other will, crudely
speaking, result in currents, $c^{(\alpha)}_i$, that are almost the
same.  In particular, vertices being part of the same (large scale)
community, are likely to be close to each other in this auxiliary
space\textsc{}~\cite{Eriksen_PRL,Simonsen-2003,Donetti2004}.  On the
other hand, vertices belonging to different communities (detected by
the mode $\alpha$) will show up with different signs for their
corresponding currents.  Such behavior is expected since the
stationary state being approached non-uniformly over the network; in
highly connected regions, like within a cluster, the stationary state
will be approached {\em faster} than in regions that are poorly
connected, as for instance between communities. If the network under
scrutiny is clustered, then often distinct, well separated, groups of
vertices, with different directions ({\it i.e.}  signs) of the
currents, will result.  Even if the network being analyzed does not
posses a community structure, the current mapping may still reveal
non-trivial topological ``secrets'' of the network~(see
Ref.~\cite{Eriksen_PRL}).

The current mapping technique consists of mapping (or projecting) the
vertices of the network onto the {\em current space}. This
$d=\alpha-1$ dimensional vector space, corresponding to the $\alpha-1$
slowest decaying modes (largest eigenvalues of ${\bm T}$ being
different from one), is constructed for vertex $i$ by associate a
point of coordinates
\begin{eqnarray}
 \label{eq:CM1}
 V_i^{(d)} &=& \left(c^{(2)}_i,c^{(3)}_i,\ldots,c^{(d+1)}_i \right).      
\end{eqnarray}

To identify communities, if any, and the vertices that belong to them,
one has to somehow cluster the points of the current
space~\cite{Eriksen_PRL,Simonsen-2003,Donetti2004}.  For a projection
space of low dimension, this can be achieved by visual inspection.  As
the dimension of the current space becomes larger, this is no longer
feasible. Instead classic clustering algorithms, like hierarchical and
optimization clustering techniques, may be
utilized~\cite{Book:Jain-1998,Book:Spaeth-1980,Book:Hartigan-1975}.
Such an approach, generalizing the ideas of the current mapping
(projection) technique of Refs.~\cite{Eriksen_PRL,Simonsen-2003}, has
recently been adapted by Donetti and Mu\~{n}oz~\cite{Donetti2004} in a
study similar in spirit to the present one. These authors applied
various types of metrics in the clustering algorithms, and found the
angular metric to perform the best.

Herein, however, we will adapt a conceptually much simpler (and more
pedagogical) approach that directly utilize the difference in signs of
the currents. The starting point of the algorithm ($\alpha=2$) is to
assign vertices of different signs for $c^{(2)}_i$ to different
partitions.\footnote{In general ${\bm c}^{(\alpha)}$ are the
  eigenvectors of ${\bm T}^\dagger$ (see
  Eq.~(\ref{eq:master-eq-vector})) corresponding to the eigenvalues
  $\lambda^{(\alpha)}$, or one may calculate them from the
  eigenvectors ${\bm \rho}^{(\alpha)}$, of ${\bm D}$.}  As the
dimension of the projection space is increased by one, a partition
from the previous step ($\alpha$) is further sub-divided if its
members correspond to different signs for the ``new'' current
$c^{(\alpha+1)}_i$. This will define a set of new {\em potential}
partitions and the modularity $Q$ (to be defined in
Eq.~(\ref{eq:modularity}) below) will be used to chose among them to
obtain the optimal partition for a given $\alpha$. A new partitioning
is only accepted if it increases the modularity as compared to the
best value obtained previously.  So for each $\alpha$, there exists an
optimal partitioning of modularity $Q^{(\alpha)}$.  In this way the
dimension of the current projection space is increased till the
modularity (and therefore the optimal partitioning) do not change any
longer with $\alpha$.  Hence, this simple clustering method is a {\em
  top-down} approach in contrast to many of the other known methods
that can be characterized as being bottom-up.

For large networks suspected to show a rich community structure, this
simple and pedagogical algorithm is, however, not optimal due to
computational cost being high when the number of communities is large.
In such cases, faster more sophisticated and complex clustering
algorithms should be
applied~\cite{Donetti2004,Book:Jain-1998,Book:Spaeth-1980,Book:Hartigan-1975}.
On the other hand for networks with limited number of communities it
performs more than adequately. It is conceptually easy to follow and
has therefore been adapted here.  Moreover, it demonstrates that the
current mapping technique does not rely on a sophisticated clustering
algorithm.

To qualitatively measure the degree of clustering for a given
partitioning of a network, the concept of {\em modularity} has
recently been introduced~\cite{Newman2003,Newman2004,Newman2004-A}. It
can be defined, for a given partitioning of a weighted network, as
\begin{eqnarray}
  \label{eq:modularity}
  Q &=&   \frac{1}{\mathcal W} \sum_{ij}\left( W_{ij} -\frac{w_iw_j}{\mathcal W} \right) \delta_{\kappa_i \kappa_j},
\end{eqnarray}
where ${\mathcal W}=\sum_{ij}W_{ij}$ is the total ``directed'' weight
of the graph,\footnote{For an undirected, unweighted network
  ${\mathcal W}$ is equal to two times the number of edges.}
$w_i=\sum_{j}W_{ji}$ the weight of outgoing edges from vertex $i$, and
$\kappa_i$ denotes the community to which vertex $i$ is assigned.

\section{Application}
\label{Sect:Results}

In this section we will present some real-world examples of the
application of the concept of diffusion to the investigation of the
topology of networks. The chosen examples correspond to networks of
both know and unknown topology, as well as being small to moderate in
size.

\subsection{Zachary Karate club network}
\label{Sect:Zachary}

\begin{figure}[btp]
  \centering
    \includegraphics*[width=13cm]{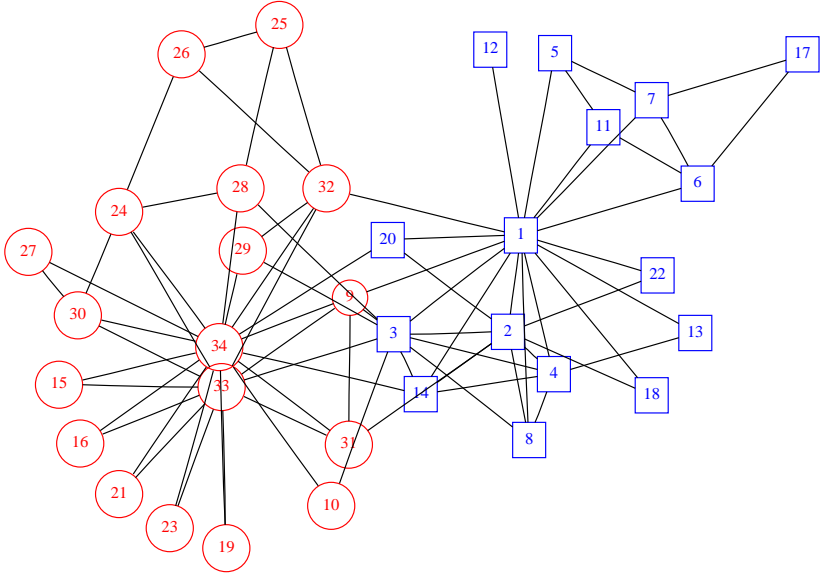}
    \caption{Zachary's friendship network~\protect\cite{Zachary} of
      the ``troubled'' karate club consisting of $34$ vertices and
      $78$ edges.  Here open squares and circles are used to denote
      the supporters, in the ongoing conflict, of the trainer (node
      $1$) and administrator (node $34$), respectively.}
  \label{Fig:ZacharyNetwork}
\end{figure}

A classic real-world network of known community structure is the
social network known as the {\em karate club network}. It has been
considered recently in a number of
studies~\cite{Girvan2002,Newman2004,Newman2004-A,Donetti2004}.
Sociologist Wayne Zachary studied in the early 1970s the relations
among the members of a karate club at an American
university~\cite{Zachary,Download}.  During the study period, it
happened by chance, that the club went through a turbulent period.  A
controversy between the club's administrator and its trainer over the
question of raising clubs fees, finally resulted in it breaking apart.
During the two years period, Zachary quantified the social ties
between the members of the club on a scale from 1 (lowest) to 5
(highest). It is the resulting weighted network that we will consider
here~\cite{Download}.  The network is depicted in
Fig.~\ref{Fig:ZacharyNetwork}, where circles and squares are used to
indicate the original partitioning obtained by Zachary. Notice that
these two communities are center around the trainer (vertex 1) and the
other around the administrator (vertex 34).
%----------------------------------------------------------
%----------------------------------------------------------

\begin{figure}[tbp]
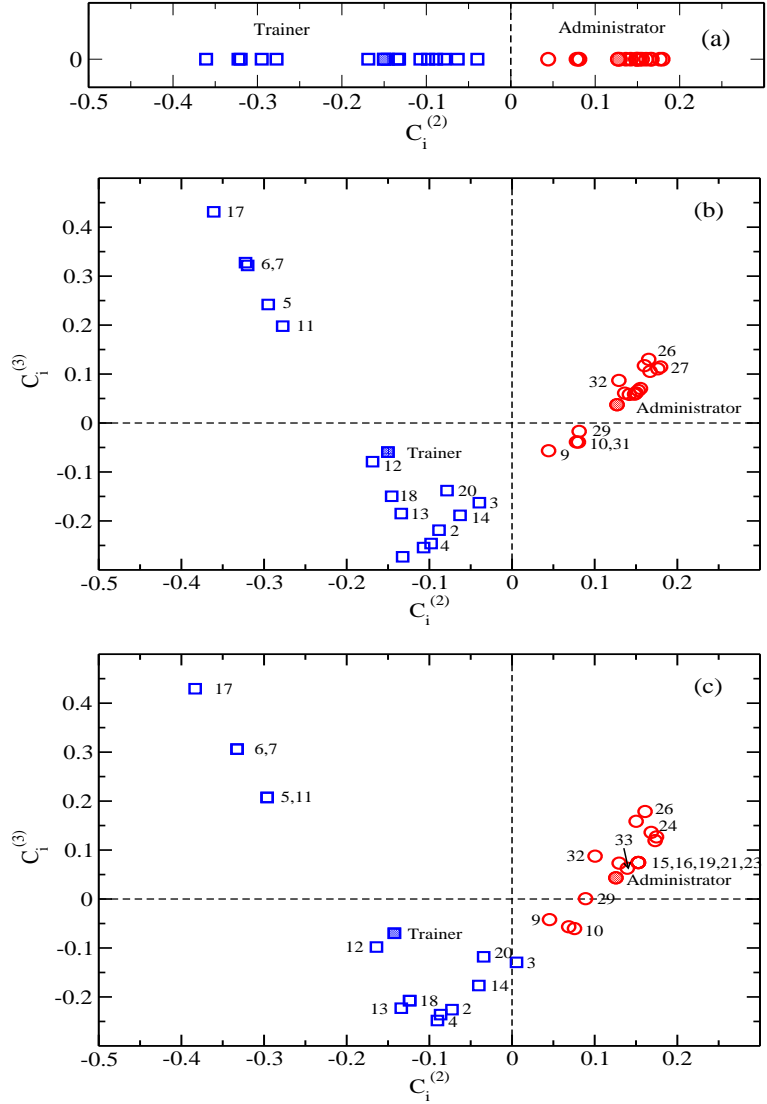

  \centering
    \qquad \includegraphics*[width=9.3cm,height=2cm]{Zacharay-weighted-c2-separation}\\*[0.3cm]
    \includegraphics*[width=10cm,height=6cm]{Zacharay-weighted-currents}\\*[0.3cm]
    \includegraphics*[width=10cm,height=6cm]{Zacharay-unweighted-currents}
    \caption{The lowest order current projections of the Zachary
      network. The (a) $c^{(2)}_i$ and (b) $c^{(2)}_ic^{(3)}_i$
      current projections of the weighted network, and (c) the
      $c^{(2)}_ic^{(3)}_i$ mapping for the corresponding unweighted
      network (result taken from Ref.~\protect\cite{Simonsen-2003}).}
  \label{Fig:Zachary-analyzed}
\end{figure}

A current mapping, based on Zachary's tie data, will now be conducted
and the results of such an analysis compared against the known
structure of the network.  Fig.~\ref{Fig:Zachary-analyzed}(a) shows
the $1$-dimensional projection of the network for the slowest decaying
($\alpha=2$) mode\footnote{Recall that $\alpha=1$ corresponds to the
  stationary state, and is thus of no interest to us in the present
  context.}.  As a guide to the eye, we have here labeled the vertices
according to the convention used in Fig.~\ref{Fig:ZacharyNetwork}, but
it should be stressed that this information has {\it not} been used
during the analysis.  Fig.~\ref{Fig:Zachary-analyzed}(a) shows a
striking division of the vertices into two groups corresponding to
positive and negative values of $c^{(2)}_i$.\footnote{Notice that the
  signs (and values) of the currents are not absolute. A
  multiplication of the eigenvector ${\bm c}^{(2)}$ by a constant may
  result in different values and signs for the currents. However,
  independent of the normalization, the relative signs of the elements
  would remain unchanged.}  This division is {\em fully} consistent
with the original classification made by Zachary.  Hence, the slowest
decaying diffusive mode $\alpha=2$ of the karate club network can be
associated with the trainer--administrator separation.  The
modularities corresponding to this division are $Q^{(2)}=0.404$ and
$Q_A^{(2)}=0.371$, where $Q_A$ refers to the modularity using the
unweighted adjacency matrix, but the same partitioning, for its
calculation\footnote{We prefer to give both these modularities for
  comparison since many authors only give $Q_A$. However, our
  partitioning was obtained using the weighted network.}.

%-------------------------------------------------------------
%-------------------------------------------------------------

%-------------------------------------------------------------

\begin{figure}[t]
  \centering
  \includegraphics*[width=10cm,height=6cm]{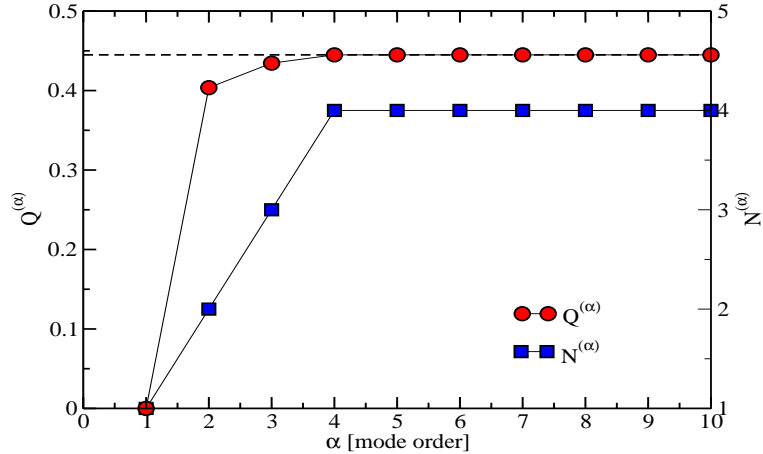}
  \caption{The modularity $Q^{(\alpha)}$ (left axis) and number of
    communities $N^{(\alpha)}$ (right axis) {\it vs} relaxation mode
    $\alpha$ for the Zachary network of
    Fig.~\protect\ref{Fig:ZacharyNetwork}. It is observed that after
    mode $\alpha=4$ no more communities are identified. The
    partitioning into four corresponds to a modularity of $Q=0.445$
    (weighted network) and $Q_A=0.420$ (unweighted network). The
    optimal value of $Q$ is indicated as a dashed horizontal line in
    the figure.}
  \label{Fig:Zachary-Final}
  \end{figure}

%-------------------------------------------------------------

%---------------------------------------------------------
\begin{figure}[bp]
  \centering
  \includegraphics*[width=12cm,height=10cm]{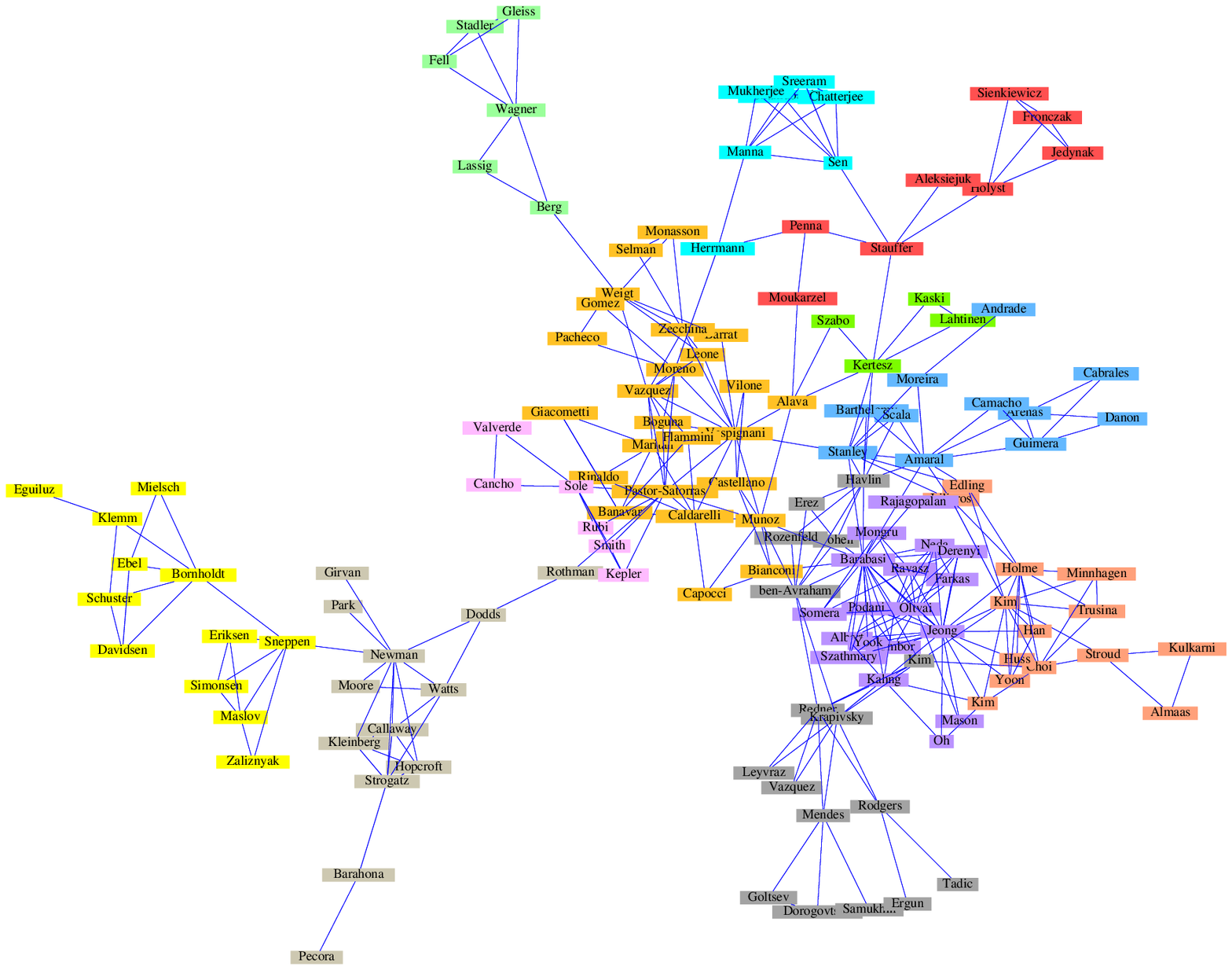}\\*[0.5cm]
  %\flushleft(b)\\
  \includegraphics*[width=10cm,height=6cm]{ScienceCollaboration-Mapping}
  \begin{picture}(0,0)(0,0)
        \put(-240,400){\makebox(0,0)[b]{$(a)$}}
        \put(-240,130){\makebox(0,0)[b]{$(b)$}}
     \end{picture}
  \centering
  \caption{A ``network'' scientists collaboration network (see
    Refs.~\protect\cite{Newman2004,Park2003} for details).  (a) The
    largest component, consists of $N=145$ vertices, of this network.
    Figure after Ref.~\cite{Newman2004}, where colors correspond to
    communities found there. (b) Same as
    Fig.~\protect\ref{Fig:Zachary-Final}, but for the network of
    Fig.~\protect\ref{Fig:ScienceCollaboration}(a).  The optimal
    partitioning is found for $14$ communities characterized by the
    modularities $Q=0.78$ and $Q_A=0.70$.  }
  \label{Fig:ScienceCollaboration}
\end{figure}
%---------------------------------------------------------

Fig.~\ref{Fig:Zachary-analyzed}(b) presents the results of performing
a $2$-dimensional current mapping of the network (modes $\alpha=2,3$).
The results suggest that the communities associated with the trainer
and administrator may be further sub-divided.  In particular, the
members $\{5,6,7,11,17\}$ are well separated from the rest of the
supporters of the trainer with different signs for the $c^{(3)}_i$
currents.  A close inspection of the network
(Fig.~\ref{Fig:ZacharyNetwork}) reveals that these members are
connected to the rest of the network {\em only} via the trainer. They
may therefore serve as good candidates for forming a trainer
sub-community.  The supporters of the administrator do also map to
$c^{(3)}_i$-currents of different signs. However, in this case, the
currents are more clustered around $c^{(3)}_i=0$ and no striking
separation between them exist. It is therefore not clear that this
separation can be attributed to a administrator sub-community. This
is, indeed, confirmed by investigating the values of the modularity of
the possible divisions.  Based on the $2$-dimensional current space, a
division into three community is optimal ($Q^{(3)}=0.435$); an
administrator community, and two communities where one consists of
members $\{5,6,7,11,17\}$, while the other one consists of the
remaining supporters of the trainer. Insisting on four communities
corresponding to the vertices located in each of the quadrants of the
$2$-dimensional current plot (Fig.~\ref{Fig:Zachary-analyzed}(b)),
would have given a modularity of $0.423$. %%%% $0.385$
This is smaller than $Q^{(3)}$ and this latter partitioning was
therefore rejected compared to the chosen one.  It is interesting to
observe that if one had based the analysis on the unweighted
network~\cite{Simonsen-2003}, the results would have been rather
similar (Fig.~\ref{Fig:Zachary-analyzed}(c)), but vertex $3$, for
instance, would not have been correctly identified\footnote{The same
  vertex, using unweighted data, was also classified incorrectly by
  one of the methods of {\it e.g.} Ref.~\protect\cite{Newman2004}.},
and there would have been more ``degeneracy'' among the current
values.

Increasing the dimension of the projection space will introduce new
potential partitions that may be accepted or rejected. The results of
gradually increasing the dimension of the projection space are
depicted in Fig.~\ref{Fig:Zachary-Final}.  Therefrom it is observed
that the optimal partitioning of the network, according to our
algorithm, is into $4$ communities that correspond to a modularity of
$Q=Q^{(4)}=0.445$ and $Q_A=Q_A^{(4)}=0.420$. Adding new modes beyond
$\alpha=4$ will not improve the partitioning. The members of the last
community, not given above, are $\{24, 25, 26, 28, 29, 32\}$.
For the same network, four communities was also reported by Newman and
Girvan~\cite{Newman2004}.  However, their communities (for the best
partitioning) were put together a little differently resulting in a
slightly lower modularity than the one reported here. Donetti and
Mu\~{n}oz~\cite{Donetti2004}, on the other hand, identified the same
communities as we did, but in addition, they had a single vertex
community (vertex 12). In effect, this difference resulted in a slight
decrease in the modularity compared to the results reported here. For
the karate club network, the partitioning given herein, results in, to
the best of our knowledge, the highest modularity values reported for
this network.

\subsection{Scientific collaboration network}

The network under scrutiny in this subsection is a collaboration
network of scientists that have published work together.  The data set
originates from Park and Newman~\cite{Park2003} and was later
restudied in Ref.~\cite{Newman2004}. The network was constructed by
taking an initial list of ``network'' scientists (actually those
appearing in the reference list of Ref.~\cite{Newman2003}) and
cross-reference those names against the physics e-print archive
arxiv.org in search for joint publications.  If, at least, one joint
work was found, an edge was created between these two scientists. Its
weight depended on the number of joint publications as well as the
number of co-authors taking part in the joint work.  Please consult
Ref.~\cite{Park2003} for further details regarding this network.  The
largest component of the resulting network is presented in
Fig.~\ref{Fig:ScienceCollaboration}(a).  This component consists of
$N=140$ scientists with the present author being among them. This
network component was recently analyzed by Newman and
Girvan~\cite{Newman2004} who reported an optimal partitioning (using
his method) consist of $13$ communities characterized by a modularity
of $Q_A=0.72\pm0.02$.

The findings using the current mapping clustering technique, are
summarized in Fig~\ref{Fig:ScienceCollaboration}(b).  It is seen that
the optimal number of clusters is found to be $14$.  The corresponding
modularities were $Q=0.78$ and $Q_A=0.70$, comparable to the result
reported in Ref.~~\cite{Newman2004}. We do not here intend to delve
into a detailed discussion on the networks community structure.
However, it should be add that our findings for the community
structure follow mainly the structure reported by Newman and
Girvan~(and indicated by vertex colors in
Fig.~\ref{Fig:ScienceCollaboration}(a)).

%------------------------------------------------------------

  \begin{figure}[tbh]
    \centering
    \includegraphics*[width=10cm,height=7.5cm]{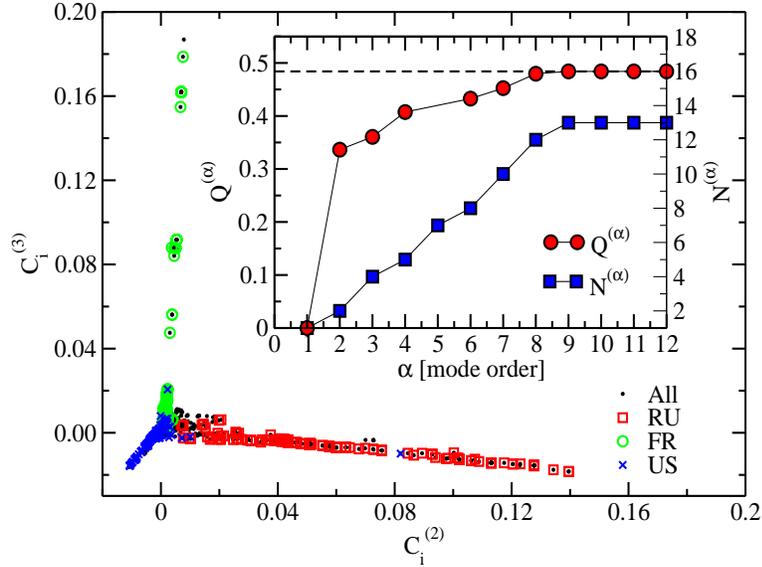}
    \caption{The two-dimensional current mapping of an Autonomous
      System (AS) network~\protect\cite{Eriksen_PRL,AS}.  The symbols
      refer to the geographical location of the AS: Russia ($\Box$),
      France ($\bigcirc$), USA ($\times$).
      %, Korea ($\bigtriangleup$).
      The inset shows the modularity and number of communities for the
      optimal partitioning of the network at a given diffusive mode.}
  \label{Fig:Internet-Currents}
  \end{figure}

%------------------------------------------------------------

\subsection{Autonomous systems}

The last example that will be considered herein is a relatively large
network where the (about $6\,500$) vertices represent so-called
autonomous systems (AS), while the edges corresponds to an entry in
the (dynamic) routing table of those devices at the time of
observation~\cite{AS}.  These networks are changing with time, and
their structures are not known in advance.

Fig.~\ref{Fig:Internet-Currents} shows the $2$-dimensional current
mapping of the networks. The star-like structure indicates that there
is a hierarchy of vertices where those located the furthest away from
the origin of the current plot are the most peripheral vertices of the
network.  Furthermore, each hierarchy corresponds roughly to the
national division of the autonomous systems network.
Fig.~\ref{Fig:Internet-Currents} shows that the three legs of the
star-structure correspond to Russia, the US and France. For the
AS-network we identified $13$ communities resulting in a modularity of
about one-half.

\section{Conclusions}
\label{Sect:Conclusions}

We have considered random walks on weighted networks. This auxiliary
network process is used to obtain information on the large scale
topological structure of the underlying network. This is done by
projecting the nodes of the network onto a low dimensional current
space. In this space, vertices that are connected to one another are
likely to appear close to each other. This is a consequence of the
relaxation towards the stationary state being non-uniform; it is
fastest in well connected regions, therefore quickly reaching a
quasi-stationary state here, and slow between poorly connected
regions.  It was found that the weights of the edges of the network
may be important to take into consideration in order to reveal the
correct underlying topology.  Furthermore, this work explicitly
demonstrates that the concept of diffusion, or random walks, is a
powerful tool that can be applied successfully to problems where no
natural connection to diffusion exists.

%---------------------------------------

\section*{Acknowledgment}

The author would like to thank K.A.\ Eriksen, K.\ Sneppen, S.\ Maslov,
S.\ Bornholdt, and M.\ H{\"o}rnquist for numerous fruitful discussions
and comments on topics related to this work.  Furthermore, the author
thanks M.E.J.\ Newman for providing the data of the science
collaboration network analyzed in Sec.~\ref{Sect:Results}.

% --------------------------------------------------------------------
% BIBLIOGRAPHY
% --------------------------------------------------------------------


\begin{thebibliography}{99}
\bibitem{Einstein1905}
A.\ Einstein, Ann. Phys. 17 (1905) 549.

\bibitem{Einstein1956}
A.\ Einstein, {\sl Investigations on the Theory of Brownian Movement}
(Dover, New York, 1956).

\bibitem{DiffusionTheory}
B.D.\ Hughes, {\sl Random Walks and Random Environments}, Vol. 1:
Random Walks,  (Oxford University Press, New York, 1995). 

\bibitem{Albert2002}        
R.\ Albert and A.-L.\ Barabasi, Rev.\ Mod.\ Phys.\ 74 (2002) 47. 

\bibitem{Newman2003}
M.\ Newman, SIAM Rev. 45 (2003) 167. 

\bibitem{Eriksen_PRL}
 K.\ A.\ Eriksen, I.\ Simonsen, S.\ Maslov, and K.\ Sneppen,
 Phys.\ Rev.\ Lett.\  90 (2003)  148701. 

\bibitem{Simonsen-2003}
I.\ Simonsen, K.\ A.\ Eriksen, S.\ Maslov, and K.\ Sneppen,
Physica A 336, (2003) 167. 

\bibitem{Girvan2002}
M.\ Girvan and M.\ E.\ J.\ Newman,
Proc. Natl. Acad. Sci. USA  99 (2002) 7821. 

\bibitem{Newman2004}
M.\ E.\ J.\ Newman, and M.\ Girvan, 
Phys. Rev. E 69 (2004) 026113. 

\bibitem{Donetti2004}
L.\ Donetti and M. A. Mu\~{n}oz, J. Stat. Mech.: Theor. Exp. (2004) P10012. 

\bibitem{Almaas2004}
 E.\ Almaas  P.L.\ Krapivsky and S.\ Redner, 
Statistics of Weighted Networks, to appear Phys.\ Rev.\ E.

\bibitem{Newman2004-A}
M.\ E.\ J.\ Newman, 
Phys. Rev. E  70 (2004) 056131.

\bibitem{Meyer-2000}
 C.D.\ Meyer, {\sl Matrix analysis and applied linear algebra}, SIAM, 2000. 

 

\bibitem{Book:Jain-1998}
   A.K. Jain and R.C. Dubes, {\sl Algorithms for clustering of data}
   (Englewood Cliffs. New Jersey, Prentice-Hall, 1988).
 

\bibitem{Book:Spaeth-1980}
  H. Spaeth, {\sl Cluster Analysis Algorithms for Data Reduction and
    Classification of Objects} Ellis Horwood 1980).
 

\bibitem{Book:Hartigan-1975}
   J.\ Hartigan, {\sl Clustering algorithms} (New York, Wiley, 1975).


\bibitem{Zachary}
W.\ W.\  Zachary,
J. Anthropol. Res. 33 (1977) 452473. 

\bibitem{Download} The Zachary network can be downloaded from : \newline
  http://vlado.fmf.uni-lj.si/pub/networks/data/UciNet/zachary.dat. 

\bibitem{Park2003}
J.\ Park and M.E.J.\ Newman, Phys. Rev. E 68 (2003) 026112. 

\bibitem{AS}
The data set can be obtained from  http://moat.nlanr.net/AS/. 

\end{thebibliography}
\end{document}